\documentclass[aip,pof,preprint,showpacs,floatfix,superscriptaddress]{revtex4-1}

\usepackage{amsmath}
\usepackage{amsfonts}
\usepackage{amsthm}
\usepackage{amssymb}
\usepackage{graphicx}
\usepackage{mathrsfs}
\usepackage{latexsym}
\usepackage{graphics}
\usepackage{color}
\usepackage{ulem}
\usepackage[english]{babel}
\usepackage{fancyhdr}

\newcommand{\be}{\begin{equation}}
\newcommand{\ee}{\end{equation}}
\newcommand{\bse}{\begin{subequations}}
\newcommand{\ese}{\end{subequations}}
\newcommand{\dst}{$\tilde{d}_s$}
\newcommand{\Hzerot}{$\tilde{H}_0$}
\makeatletter
\newcommand{\vast}{\bBigg@{2.5}}
\makeatother

\begin{document}

\title{Surface wave dynamics in orbital shaken cylindrical containers}
\author{M. Reclari}
\author{M. Dreyer}
\affiliation{EPFL, Laboratoire des Machines Hydrauliques, 1007 Lausanne, Switzerland}
\author{S. Tissot}
\affiliation{EPFL, Laboratory of Cellular Biotechnology, 1015 Lausanne, Switzerland}
\author{D. Obreschkow}
\affiliation{International Centre for Radio Astronomy Research (ICRAR), M468, University of Western Australia, 35 Stirling Hwy, Crawley, WA 6009, Australia }
\author{F. M. Wurm}
\affiliation{EPFL, Laboratory of Cellular Biotechnology, 1015 Lausanne, Switzerland}
\author{M. Farhat}
\affiliation{EPFL, Laboratoire des Machines Hydrauliques, 1007 Lausanne, Switzerland}

\begin{abstract}

Be it to aerate a glass of wine before tasting, to accelerate a chemical reaction or to cultivate cells in suspension, the ``swirling'' (or orbital shaking) of a container ensures good mixing and gas exchange in an efficient and simple way. Despite being used in a large range of applications this intuitive motion is far from being understood and presents a richness of patterns and behaviors which has not yet been reported. The present research charts the evolution of the waves with the operating parameters identifying a large variety of patterns, ranging from single and multiple crested waves to breaking waves. Free surface and velocity fields measurements are compared to a potential sloshing model, highlighting the existence of various flow regimes. Our research assesses the importance of the modal response of the shaken liquids, laying the foundations for a rigorous mixing optimization of the orbital agitation in its applications.\vspace{3em}\\*
\footnotesize\textit{Copyright (2014) American Institute of Physics. This article may be downloaded for personal use only. Any other use requires prior permission of the author and the American Institute of Physics.}\vspace{3em}\\*
\footnotesize\textit{The following article appeared in Physics of Fluids \textbf{26}, 052104 (2014) and may be found at \url{http://scitation.aip.org/content/aip/journal/pof2/26/5/10.1063/1.4874612}.}
\end{abstract}
\maketitle

\section{Introduction}
The ``orbital shaking'' is the displacement of a container, maintaining a fixed orientation with respect to an inertial frame of reference, along a circular trajectory at a constant angular velocity. This motion is used in many applications, ranging from wine swirling for tasting purposes to biological and chemical industrial applications, e.g.~bacterial \cite{McDaniel1969} and more recently cellular cultures \cite{WurmNatureBT2004}. Cells are cultivated in suspensions of liquid medium, where they multiply and produce the required protein. The motion of the liquid guarantees the suspension and homogenizes the concentration of cells consumables (oxygen and nutrient) and products (proteins and carbon dioxide). Although at large scale this is usually achieved using stirred tanks, where the agitation is provided by an impeller and the aeration by oxygen spargers from the container bottom \cite{Ullmann2000}, it was found that bursting of bubbles at the free surface \cite{HandaCorrigan1989} and excessive shear stresses\cite{Kretzmer1991,Papoutsakis1991} may damage the cells. These reasons kindled the interest in the use of large scale orbital shaken devices up to thousand liters \cite{LiuHong2001,MullerWurm2007,DeJesusWurm2004}. Thus, a large number of researches have been performed on the gas exchange \cite{Buchs2001,Muller2005,MaierBuchs2004,Zhang2009}, mixing \cite{Micheletti2006,TanBuchs2011,Tissot2010,Tissot2011} and volumetric power consumption\cite{Buchs2000} of the bioreactors. In some cases, correlations have been proposed in order to quantify the variation of those quantities during scale-up of the culture size. Although few studies focused on the measurement of the velocity fields \cite{Walker2003,Weheliye2013}, no significant effort has been dedicated so far to the understanding of the motion of the liquid carrying the cells, and of its influence on the scaling of the cultures.

On the other hand, the physics of free surface liquids in laterally shaken containers has been extensively studied, especially from the analytical point of view, with potential flows theory (inviscid, incompressible and irrotational) \cite{Ibrahim2005}. Early works focused on sloshing of spacecraft propellent in cylindrical \cite{Lomen1965a,Lomen1965b}, conical \cite{Bauer1988} or rectangular tanks \cite{Bauer1964}, while more recent investigations involved the transport of liquids in naval carriers \cite{Faltinsen2003, Lee2007}. Moreover, swirling and unsteady flow regimes have been observed in linear forced sloshing at shaking frequencies close to the first natural frequency \cite{Hutton1963, Abramson1966}. Extensive study determined the stability regions of those flows\cite{Miles1984Resonantly, Faltinsen2003, Royon-Lebeaud2007}, while the persistence of swirling regime due to orbital shaking has also been investigated \cite{Faller2001}. Surprisingly, the hydrodynamics generated by orbital excitation and its application for mixing and oxygenation purposes received little attention so far. 
This work bridges the gap between the biological and hydrodynamic fields of investigation, explicitly establishing a potential model for the case of orbital shaken liquids. We also conducted an extensive experimental survey of the free surface and liquid motion to asses the limits of validity of the potential model. 

\section{Analytical potential model}
\begin{figure}[htb] 
	\centering
	\includegraphics{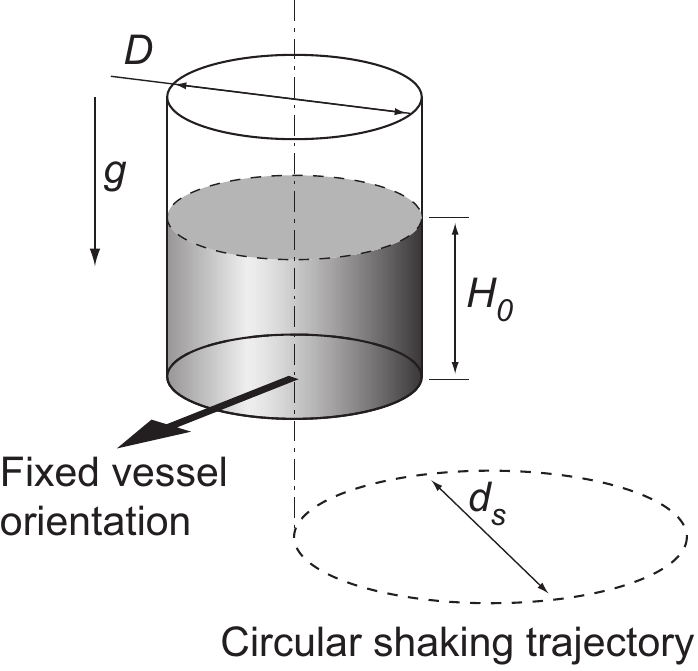} 
	\caption{Schematic illustration of the operating parameters of the shaking configurations.}
	\label{Fig_SetupOrbShak} 
\end{figure}

We consider the traditional case of a potential flow (inviscid, irrotational and incompressible) within a circular cylinder with upright wall. The velocity is therefore described as the gradient of a potential $\Phi$, solution of the Laplace's equation with impermeability conditions at the container walls, and kinematic and dynamic boundary conditions at the free surface. Following the resolution method used for linear forcing\cite{Ibrahim2005} we solve this equations for orbital shaking. The origin of the cylindrical reference system is at the container revolution axis, at the unperturbed height of the liquid. The parameters characterizing each shaking configuration are the inner diameter of the cylindrical container $D$, the diameter of the circular shaking trajectory $d_s$, the height of the liquid at rest $H_0$ and the shaking frequency $\Omega$ (Fig.~\ref{Fig_SetupOrbShak}).
The orbital shaking motion may be described as a combination of two sinusoidal translations with $\pi/2$ phase shift, which expressed in terms of velocity leads to the following equations for the motion of the wall:
\be
		\dot{\mathbf{X}}_0(t)= \left\{
		\begin{array} {l}
			\medskip -\frac{d_s \Omega}{2} \sin(\Omega t -\theta) \mathbf{e}_r \\
			\frac{d_s \Omega}{2} \cos(\Omega t-\theta) \mathbf{e}_{\theta}. 
		\end{array} \right. \label{eq_Slosh_SloshForcingVelCyl}
\ee
It is usual to separate the potential function $\Phi$ into a liquid motion component $\bar{\Phi}$ and a container motion component $\Phi_0$. Moreover, the free surface boundary conditions are linearized and the radius $r$, appearing in the forcing terms, is expanded in a Bessel-Fourier series. The solution of this set of equations has the form of a series of modes in the tangential and radial directions. They are usually identified by the number of nodal diameters $m$ and nodal circles $n$ respectively. The orbital shaking is likely to excite non-axysimmetric modes: we consider therefore the linear solution formed by the $m$=1 modes, having the following form:
\begin{align}
	& \bar{\Phi}(r,\theta,z,t)=-\frac{d_s}{2}\Omega 
	\left[\sin{\Omega t}\cos{\theta} - \cos{\Omega t}\sin{\theta} \right] \nonumber \\
	&\quad \times \sum_{n=1}^{\infty} \vast[ \frac{D}{(\varepsilon_{1n}^2-1)} \frac{\Omega^2}{(\omega_{1n}^2-\Omega^2)} \frac{J_1(2\varepsilon_{1n}r/D)}{J_1(\varepsilon_{1n})} \label{eq_Slosh_ResultTildePhi}\\
	& \quad \quad \quad \quad \frac{\cosh(2\varepsilon_{1n}(z+H_0)/D)}{\cosh(2\varepsilon_{1n}H_0/D)} \vast]. \nonumber
\end{align}
where $J_1$ is the Bessel's function of the first kind and first order, $\varepsilon_{1n}$ are the $n$ roots of the derivative of $J_1$, and $\omega_{1n}$ are the natural frequencies calculated according to:
\be
	\omega_{1n}^2 = \frac{2g \varepsilon_{1n}}{D}\tanh{ \left( \frac{2\varepsilon_{1n}H_0}{D}\right)}. \label{eq_Slosh_EigenFrequencies}
\ee
The corresponding free surface elevation $\xi(r,\theta,t)$ has the following form:
\begin{align}
	& \xi(r,\theta,t)=\frac{d_s \Omega^2}{2 g} \cos(\Omega t-\theta) \nonumber \\
	& \times \left\{r+\sum_{n=1}^{\infty} \left[ \frac{D}{(\varepsilon_{1n}^2-1)} \frac{\Omega^2}{(\omega_{1n}^2-\Omega^2)} \frac{J_1(2\varepsilon_{1n}r/D)}{J_1(\varepsilon_{1n})}  \right] \right\}. \label{eq_Slosh_FreeSurf} 
\end{align}
Equation \ref{eq_Slosh_FreeSurf} may be rewritten in a scale independent way by introducing the following dimensionless quantities: $\tilde{r}=r/D$, $\tilde{z}=z/D$ and $\tilde{\xi}=\xi/D$. We define the Froude number $Fr^2=d_s \Omega^2/g$. Accordingly, we define the following dimensionless operating parameters: $\tilde{d}_s=d_s/D$ and $\tilde{H}_0=H_0/D$. Those dimensionless parameters ensure therefore the hydrodynamic similarity between different scales. The free surface height is hence defined as:
\begin{align}  
	& \tilde{\xi}(\tilde{r},\theta,t)=\frac{Fr^2}{2}\cos(\theta-\Omega t) \nonumber \\
	& \times \left\{\tilde{r}+\sum_{n=1}^{\infty} \left[ \frac{1}{(\varepsilon_{1n}^2-1)} 
	\frac{Fr^2}{(Fr_{1n}^2-Fr^2)} \frac{J_1(2\varepsilon_{1n}\tilde{r})}{J_1(\varepsilon_{1n})}  \right] \right\} \label{eq_Slosh_adimFreeSurfModel}
\end{align}
where $Fr_{1n}$ is found according to Eq.~\ref{eq_Slosh_EigenFrequencies}:
\be
	Fr_{1n}^2=2 \varepsilon_{1n} \tilde{d}_s \tanh (2\varepsilon_{1n} \tilde{H}_0).
\ee

The weakly non-linear solutions are obtained by expanding the free surface boundary conditions. Those solutions allow a better estimation of the forces acting on the container and a better agreement with the non-linear behavior of the wave near the natural frequencies \cite{Hutton1963, Royon-Lebeaud2007}. The traditional solution for sloshing under longitudinal excitation consists in considering the (1, 1) mode as dominant, and the (0, 1) and (2, 1) as the most relevant secondary modes, having an amplitude of the order of $O(a_{11}^2)$ where $a_{11}$ is the amplitude of the dominant mode \cite{Penney1952}. The free surface boundary conditions are expanded in Taylor series around $z$=0, and only the orders $a_{11}$ and $a_{11}^2$ are kept. The solution for the free surface, obtained using a solving strategy inspired from linear forcing\cite{Abramson1966}, has the following form:
\begin{align} \label{eq_Slosh_NonLinFSSolution}
	& \xi(r,\theta,t)= A_{11} \cos(\Omega t-\theta)J_1(\lambda_{11}r) \nonumber \\
	& \quad- A_{01} J_0(\lambda_{01}r)  + A_{21} \cos(2(\Omega t-\theta)) J_2(\lambda_{21}r)
\end{align}
where the amplitudes are calculated as:
\bse \begin{align}
	& A_{11}= \frac{d_s D \Omega^2 \omega_{11}^2}{2 g (\omega_{11}^2 -\Omega^2) (\varepsilon_{11}^2-1)J(\varepsilon_{11})} \label{eq_Slosh_A11}\\
	& A_{01}=-\frac{d_s^2 \Omega^6}{8 g (\omega_{11}^2 \Omega^2)^2(\varepsilon_{11}^2-1)J(\varepsilon_{11})^2} \left[n_{01,1}- n_{01,2}\frac{\omega_{11}^2 D^2}{g^2} \right] \label{eq_Slosh_A01}\\
	& A_{21}=-\frac{d_s^2 \Omega^4 \omega_{11}^2 \omega_{21}^2 }{2 g (\omega_{11}^2 -\Omega^2)^2 (\omega_{21}^2-4\Omega^2)(\varepsilon_{11}^2-1)J(\varepsilon_{11})^2} \nonumber \\
	& \quad \cdot\left[n_{21,1} + n_{21,2} \frac{\Omega^2}{\omega_{11}^2} - n_{21,3} \frac{\omega_{11}^2 \Omega^2 D^2}{g^2}
	-n_{21,1}\frac{(\omega_{21}^2 -4\Omega^2)}{\omega_{21}^2} \right]. \label{eq_Slosh_A21}
\end{align} \ese
and the coefficients, obtained by numerical integration, are the following:
\begin{align}
	& n_{01,1}= 0.4046	\nonumber \\ 
	& n_{01,2}= 0.0298 \nonumber \\ 
	& n_{21,1}= 1.6362 \\					 
	& n_{21,2}= 0.447 \nonumber \\ 
	& n_{21,3}= 0.2631 \nonumber		 
\end{align}

\section{Experimental setup}
To obtain a smooth and steady orbital motion, with minimum shocks and jerks, we used a Kuhner Es-X shaker (420 x 420mm, 25kg of maximum load, 20 to 500 RPM), which we have modified to allow for a continuous adjustment of the shaking diameter during operation.

\begin{figure*}[tb] 
	\includegraphics{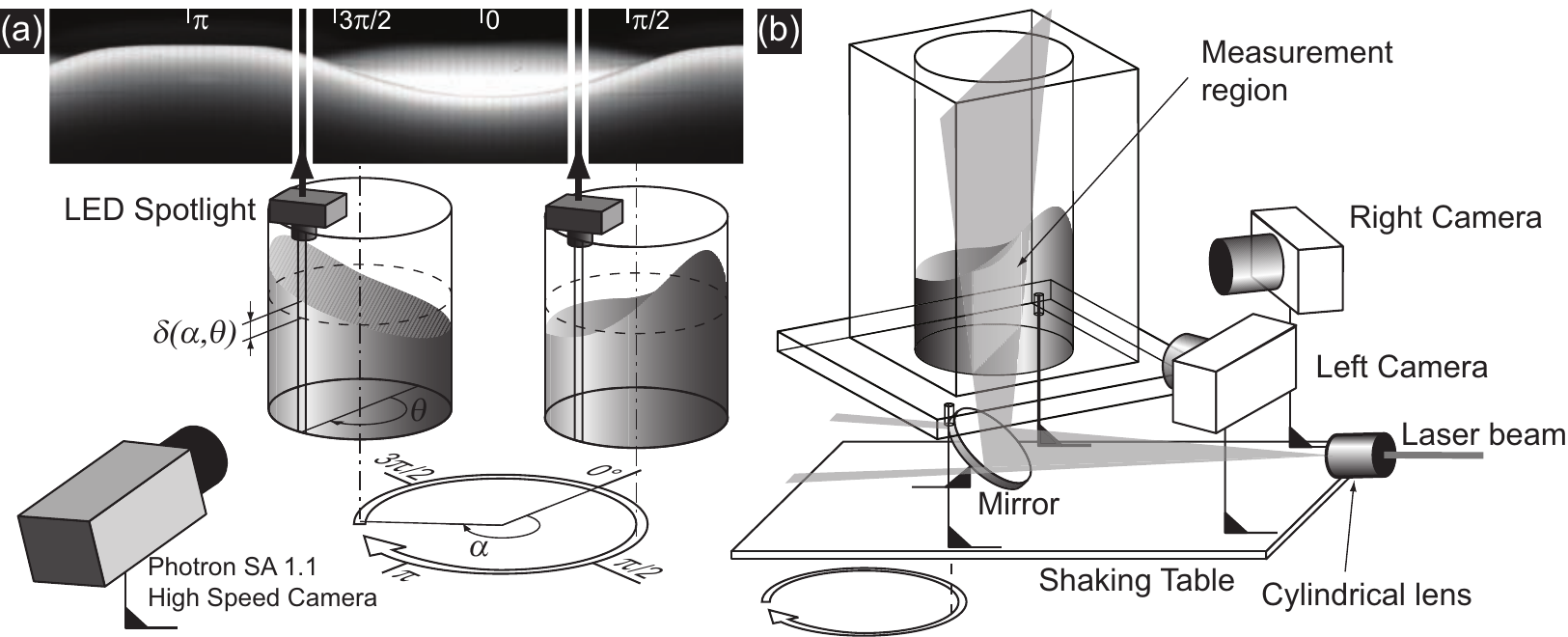} 
	\caption{\textbf{(a)} Procedure used to measure the surface displacement at the wall ($\delta(\alpha)$) and to retrieve a single image of the wave from a series of frames taken from a fixed location by the high speed camera. \textbf{(b)} Stereo PIV setup. The cameras have been equipped with Scheimpflug mounts not depicted here.}
	\label{Fig_Setups} 
\end{figure*}

We have carried out two sets of experiments. In the first one, we have measured the free surface deformation at the wall for a wide range of shaking configurations. To this end, a high speed camera (Photron SA 1.1) was used to record the motion of the free surface illuminated by a narrow beam spot light (Figure 2.a). The contrast was enhanced by adding a small amount of white dye to the liquid ($<$1\% of total volume). The periodicity of the wave motion in space and time, which is imposed by the constant agitation frequency, was exploited to retrieve the water height, along the container periphery using the high speed movies. A column of pixels was extracted from each frame and processed to estimate the water height at a fixed location of the container wall by exploiting the strong contrast between the illuminated liquid and the dark background. Setting side by side those columns of pixels gives a snapshot of the wave pattern along the wall, which also represents the time variation of the water height at a fixed angular position (Fig.~\ref{Fig_Setups}a). The frame rate and the column width (in pixels) were adjusted so that the reconstructed images have a total width of 1000 pixels for each vessel revolution. The measurements were phase averaged over 3 rotations. Since the control and data acquisition were made fully automatic, we could measure the wave height for over 6’000 operating configurations, with $D$=144 and 287 mm, $d_s$ ranging from 2 to 65mm, $H_0$ between 45 and 150mm and shaking frequencies increasing stepwise from 20 to 200 rpm.

In the second experiment, we have measured the velocity field in the liquid phase with the help of Stereoscopic Particle Image Velocimetry (SPIV) for a subset of operating conditions (Fig.~\ref{Fig_Setups}b). The setup is composed of a couple of Dantec FlowSense EO 4M cameras (resolution 2048 X 2048 pixels, 8 bits) and a Litron Dual Power 200-15 laser (pulse energy 2x200 mJ, wavelength 532 nm). The cameras and the mirror were mounted on the shaken table, while the laser source and the optics used to obtain the light sheet were kept still. The cameras were equipped with the so called Scheimpflug mounts (not depicted in the figure) to align the focal plane with the laser plane. The deformation of the image due to the presence of a non-planar wall was mitigated using a PMMA (Polymethyl-Methacrylate) container with square external and cylindrical internal walls. To avoid the effects of light scattering by bubbles, free surface and walls, the tracking polyamide particles (polyamide particles, 5 - 35$\mu$m diameter, density 1.03kg/m$^3$) have been coated with a fluorescent dye (Rhodamine B), which emits orange light at 620nm when excited at 540nm \cite{Kubin1983}. The cameras were equipped with long pass filters, which cut wavelengths shorter than 570 nm. As predicted by the potential model, and suggested by visual observations of the wave, the velocity field is constant in a reference frame rotating at the shaking frequency. Therefore, the measurements were performed on a vertical plane whose position is fixed with respect to the container. Measurements taken at different time steps within the shaking cycle give the velocity field of the entire liquid volume.

\section{Results}
\begin{figure}[htb] 
	\centering 
	\includegraphics{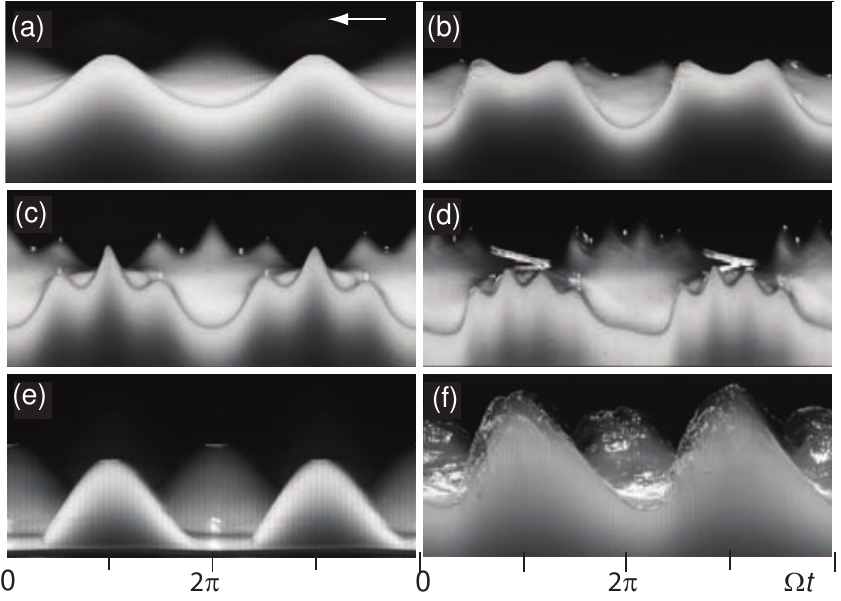} 
	\caption{Wave patterns reconstructed from high speed movies, depicting two revolutions of the vessel. All waves are travelling from right to left. \textbf{(a)} single crested wave, the most usually observed. \textbf{(b)} double crested wave. \textbf{(c)} triple crested wave. \textbf{(d)} quadruple  crested wave. \textbf{(e)} wave drying a portion of the vessel bottom. \textbf{(f)} breaking single crested wave.}
	\label{Fig_Res_WavePatterns} 
\end{figure}

The observation of the free surface shapes for a large number of shaking configurations reveals a remarkable richness of wave patterns. Although single crested waves, with one crest and one trough (Fig.~\ref{Fig_Res_WavePatterns}a) are the most common, more complex shapes featuring multiple crests and troughs are also observed: double (Fig.~\ref{Fig_Res_WavePatterns}b), triple (Fig.~\ref{Fig_Res_WavePatterns}c) and quadruple crest waves (Fig.~\ref{Fig_Res_WavePatterns}d). Under specific conditions, the wave may ``dry'' a portion of the vessel bottom (Fig.~\ref{Fig_Res_WavePatterns}e) or break (Fig.~\ref{Fig_Res_WavePatterns}f). 

\begin{figure*}[htb] 
	\centering 
	\includegraphics{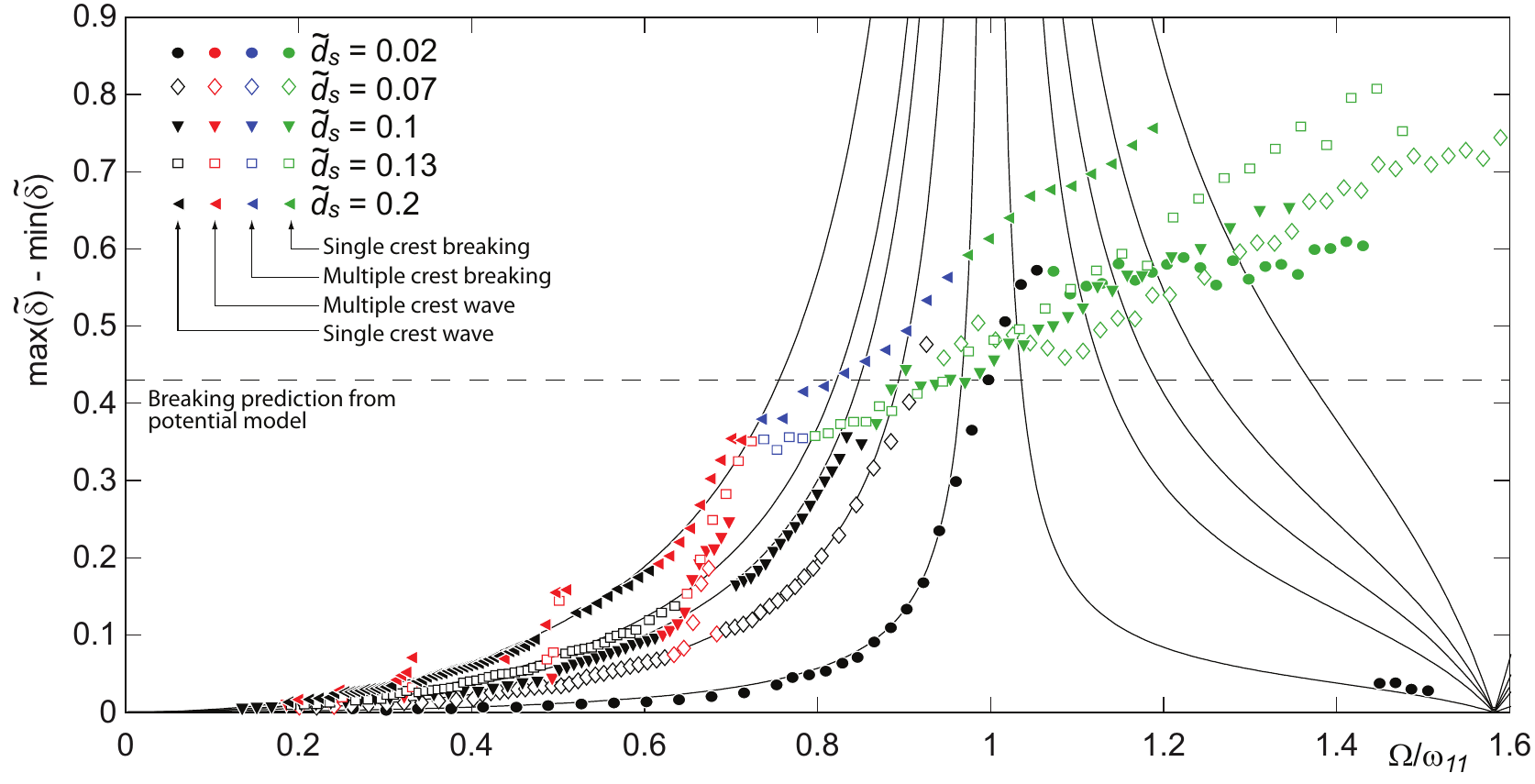} 
	\caption{Measured crest-to-trough amplitudes for \Hzerot=0.52 and five values of \dst, as a function of the shaking frequency $\Omega$ normalized by the first natural frequency $\omega_{11}$. The kind of wave is indicated by the color of the marker. The prediction of the linear solution of the potential model are depicted as solid lines, while the limiting amplitude at which waves break (Eq.~\ref{eq_res_breakingLimit}) is depicted as a dashed line.}
	\label{Fig_AmpliCompare} 
\end{figure*}

Figure \ref{Fig_AmpliCompare} shows the dimensionless crest-to-trough amplitude of waves measured at the wall ($\tilde{A}_{\delta} \equiv \text{max}(\tilde{\delta}(\theta,t))-\text{min}(\tilde{\delta}(\theta,t))$, where $\tilde{\delta}(\theta,t)=\delta(\theta,t)/D$) as a function of the shaking frequency $\Omega$ normalized by the first natural frequency $\omega_{11}$, for five values of the shaking diameter \dst. The measurements are compared to the amplitudes predicted by the linear solution of the potential model, which are computed according to Eq.~\ref{eq_Slosh_adimFreeSurfModel} and depicted as a solid line for each shaking diameter in Fig.\ref{Fig_AmpliCompare}. The color of the marker distinguishes between the single crested breaking waves (black), the waves showing multiple crests (red), multiple crested breaking (blue) and the single crested breaking/splashing (green).

Despite the strong hypothesis leading to the linear solution of the potential model, we observe that the experiments and the model agree remarkably well as long as the wave remains single crested and does not break. For example the multiple crested waves appearing at $\Omega/\omega_{11}\cong 0.5$ display amplitudes larger than predicted, but the agreement with the potential prediction is resumed as the shaking frequency is increased and the wave returns to single crested shape.

Multiple crested waves are observed at fractions of the natural frequencies, and are therefore likely to be the sub-harmonic waves of the natural modes. For example, the non-linear solution of the potential model (Eq.~\ref{eq_Slosh_NonLinFSSolution}) predicts an increase of the amplitude at a shaking frequency equivalent to half of the secondary natural frequency $\omega_{21}$ (see Eq.~\ref{eq_Slosh_A21}), leading to a double crested wave. To determine the modes and sub-harmonics contributing to the multiple crests occurrence we have decomposed the waves height measured at the wall $\delta(\theta,t)$ into a single crested part $\xi_{1n}(D/2,\theta,t)$, according to the linear prediction (i.e.~Eq.~\ref{eq_Slosh_adimFreeSurfModel}), and a secondary contribution: $\delta_{sh}(\theta,t)$, fitted (in the least squares sense) with the following generic wave expression: 
\be
	\delta_{sh}(\theta,t)=\frac{A_{sh,p}}{2}\cos \left( p ( \Omega t -\theta) +\phi_{sh,p} \right), \label{eq_SubHarmonicDef}
\ee
where $p$ is an integer, the number of crests and troughs of the wave. The fitting determines the value of $p$ best approximating the experimental data (smallest value of fitting residuals), and the corresponding amplitude $A_{sh,p}$ and phase shift $\phi_{sh,p}$. 
\begin{figure*}[htb] 
	\centering 
	\includegraphics{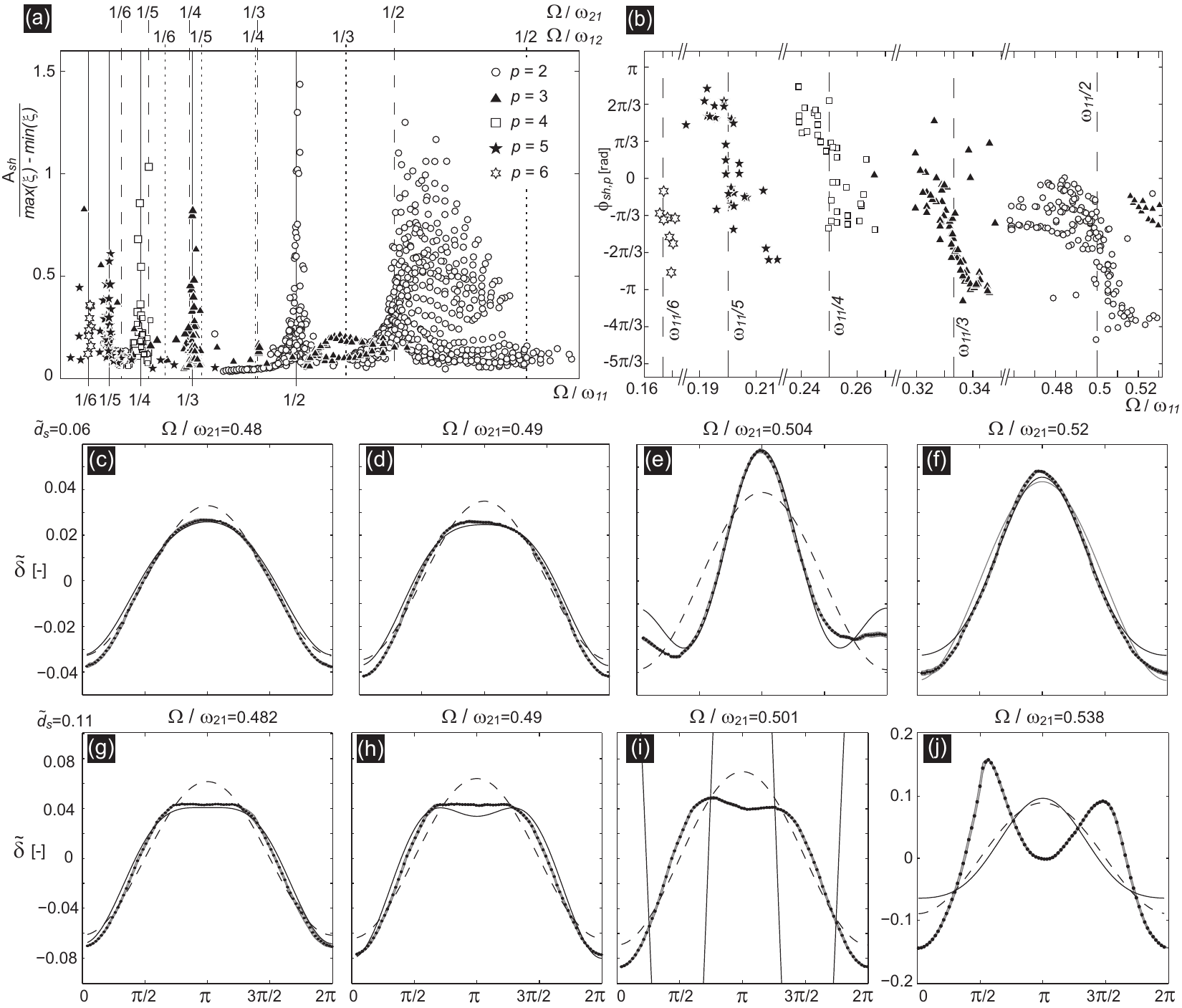} 
	\caption{\textbf{(a)} Amplitude $A_{sh,p}$ and \textbf{(b)} phase shift $\phi_{sh,p}$ of the sub-harmonics contribution $\delta_{sh}(\theta,t)$ according to the best fitting value of $p$ (see Eq.~\ref{eq_SubHarmonicDef}) normalized by the amplitude of the wave predicted by the linear solution of the potential model, as a function of the shaking frequency (scaled by the natural modes $\omega_{11}$ below the graph, by the modes $\omega_{21}$ and $\omega_{12}$ above). \Hzerot=0.52, \dst=0.01 to 0.4 in steps of 0.01. \textbf{(c)-(f)} Comparison between the dimensionless wave height measured at the wall $\tilde{\delta}(\theta)$ (depicted as black dots, while the standard deviation is shown as the gray surface) and the linear (dashed line) and non-linear (solid line) solution of the potential model, \Hzerot=0.52, \dst=0.1. \textbf{(g)-(j)} Same as \textbf{(c)-(f)}, with \Hzerot=0.52, \dst=0.2. Note that in \textbf{i} the non-linear prediction has a very large amplitude, therefore only four nearly vertical lines are visible.}
	\label{Fig_SubHarmWaves} 
\end{figure*}
The results for the amplitude are depicted in Fig.~\ref{Fig_SubHarmWaves}a, where the shaking frequencies corresponding to fractions of the natural modes are highlighted. We notice that not only the sub-harmonic of the mode $(2, 1)$ is excited, but also those of the dominant mode (1, 1) and, to a less extent, those of the (1, 2) mode. In several cases, subharmonics of two or more natural frequencies are close to each another (e.g.~$\omega_{11}/3\cong\omega_{21}/4$). We notice that the amplitude of the lowest natural frequency is dominant: at $\Omega \cong \omega_{11}/3 \cong \omega_{21}/4$ the best fitting value of $p$ is 3, and the wave is triple crested. Both sub-harmonics are nevertheless excited, although the (2, 1) with lower amplitude. We notice in Fig.~\ref{Fig_SubHarmWaves}b that the phase shift $\phi_{sh,p}$ is subject to a change by $\pi$ as the shaking frequency crosses one of the sub-harmonics frequencies, as expected in resonant systems. On the other hand, the phase of primary waves is not affected. The non-linear solutions of the potential model predicts therefore only one between several surface perturbation: the one having the largest amplitudes.

The measured occurrence of sub-harmonic waves is compared to the predictions of the non-linear model in Figures \ref{Fig_SubHarmWaves}c-j, for the first sub-harmonics of the (2, 1) mode, Eq.~\ref{eq_Slosh_NonLinFSSolution}. We observe that at low shaking diameters the non-linear model correctly predicts the amplitude of the sub-harmonic wave at shaking frequencies below and above $\omega_{21}/2$. On the other hand, as \dst~is increased and $\Omega>\omega_{21}/2,$ the wave displays a persistence of double crested shape, the sub-harmonic do not shift its phase as predicted by the potential model (Fig.~\ref{Fig_SubHarmWaves}j). The validity of the non-linear prediction is therefore limited by the shaking diameter.

\begin{figure}[tb] 
	\centering 
	\includegraphics{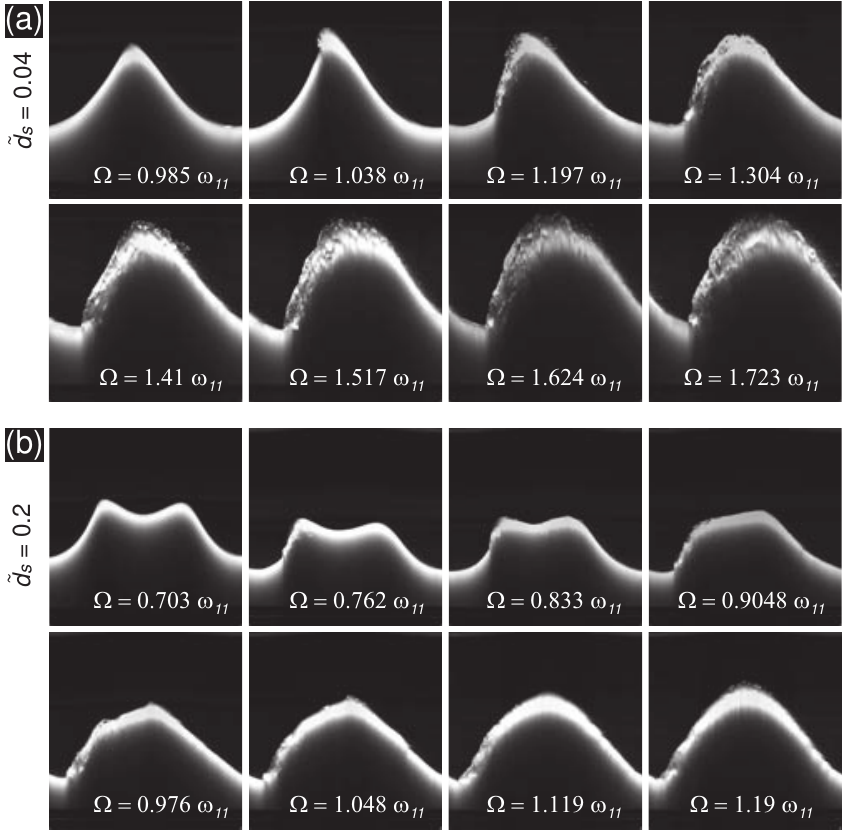} 
	\caption{Reconstructed snapshots of waves from shaking frequencies slightly below the breaking inception of breaking to larger shaking frequencies; \Hzerot=0.52. \textbf{(a)} \dst=0.04, \textbf{(b)} \dst=0.2.}
	\label{BreakingWaves} 
\end{figure}

The resolution of the gravity waves equations at orders higher than the first one demonstrates the existence of a limiting steepness, beyond which no gravity wave can exist, thus introducing a threshold for the wave breaking \cite{Stokes1880, Grant1973}. As research on wave breaking progressed, it was found that several instabilities are liable to develop before this limit is reached \cite{Longuet-Higgins1997}. For gravity waves, the limiting steepness is $H/\lambda \cong 0.1412$, where $H$ is the crest-to-trough amplitude and $\lambda$ is the wavelength \cite{Schwartz1982}. In the specific case of orbital shaken containers, $H=\max(\delta)-\min(\delta)$ and $\lambda=\pi D$, therefore the threshold amplitude is:
\be
	\max(\tilde{\delta})-\min(\tilde{\delta}) \cong 0.443. \label{eq_res_breakingLimit}
\ee
This is the upper limit at which waves should break, and it is depicted as a dashed line in Fig.~\ref{Fig_AmpliCompare}. We also observe in Fig.~\ref{Fig_AmpliCompare} a change of the wave behavior when the wave breaks, at amplitudes between 0.35 and 0.45. Both single and multiple crested breaking waves were observed. If the double crested waves generated at $\Omega=\omega_{21}/2$ persists long enough as $\Omega$ is increased, the wave breaks as double crested. According to our investigation, the discriminant shaking diameter above which waves breaks as double crested may be empirically determined as: 
\be
	\tilde{d}_{s,\omega_{21}/2}=0.1349 \cdot \tanh^2(2.5 \tilde{H}_0) \label{eq_Res_Crit_dstilde}.
\ee
Moreover, we have observed that multiple crested waves break at lower amplitudes compared to single crested ones. Indeed, the slope at the front crest of double crested waves is steeper than the one of single crested waves having similar crest-to-through amplitude. The wave breaking is initially visible only at the wave crest, in the form of a splash of water at the wall (Fig.~\ref{BreakingWaves}). In multiple crested waves the breaking appears at the steepest crest, usually the front one (Fig.~\ref{BreakingWaves}b). The breaking then becomes more generalized, and we observe spilling breakers, with entrainment of air bubbles into the liquid. As $\Omega$ is further increased the ``white water'' at the crest spreads down over the front slope and the wave loses its symmetry. The breaking starts at the wall, where the wave is steeper, and propagates to a limited extent (usually less than one third of the radius) toward the inner part of the container.

As the shaking frequency is increased above the breaking inception, the liquid motion becomes more similar to a jet impinging at the wall than a wave. The liquid is mostly at the wall, and the free surface near the container axis is depressed. Moreover, the sub-harmonic wave giving the double crests gradually disappears, and completely vanishes before reaching $\omega_{11}$. As the shaking frequency increases, the morphology of single and multiple crested waves becomes similar notwithstanding the number of crests at the breaking inception, and the amplitude increases almost linearly with the shaking frequency (as we observe in Fig.~\ref{Fig_AmpliCompare}).

\begin{figure*}[tb] 
	\centering 
	\includegraphics{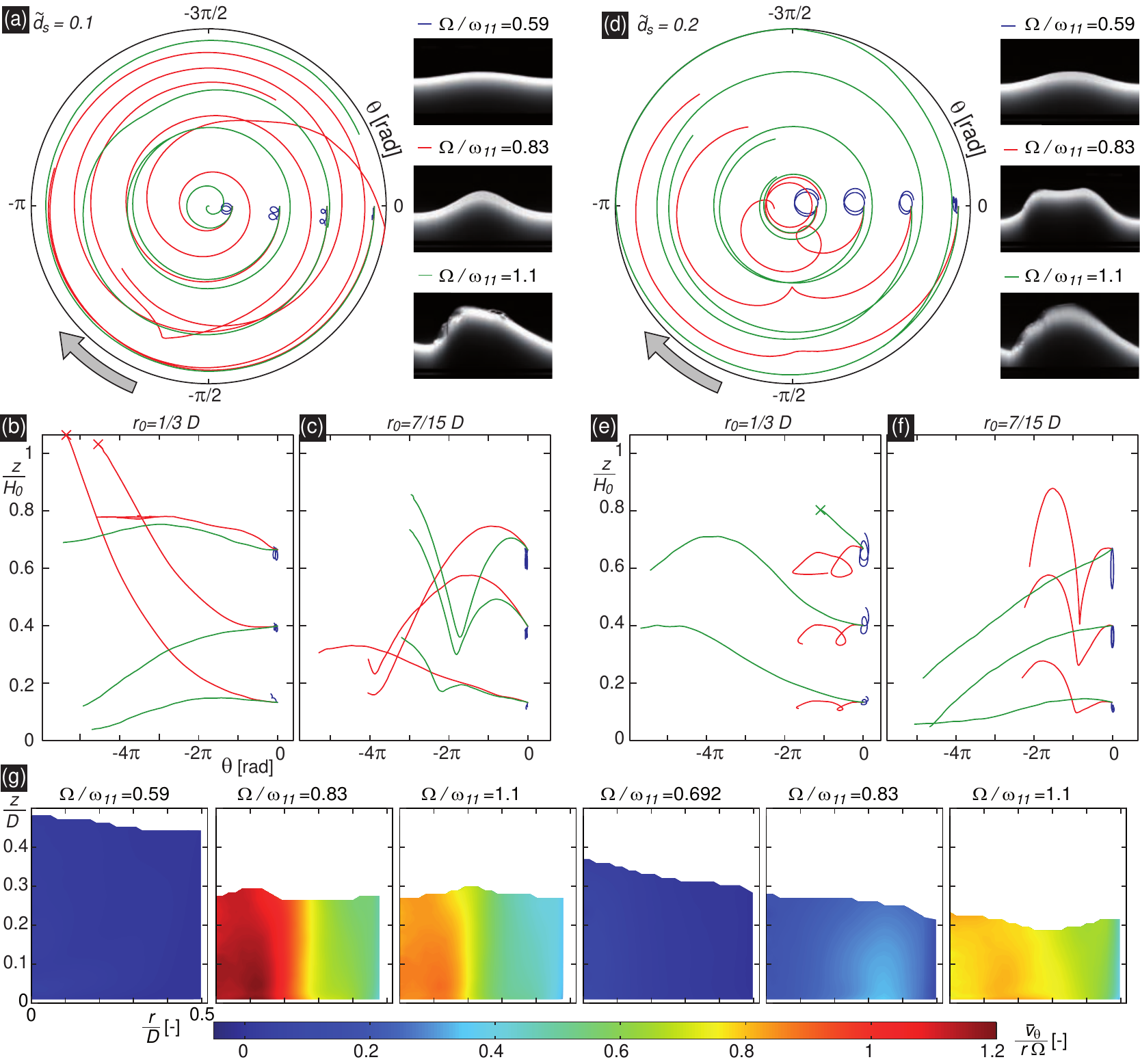} 
	\caption{(Colors online) Lagrangian trajectories of liquid particles reconstructed from PIV measurements, for \Hzerot=0.5, \dst=0.1 (\textbf{(a)} to \textbf{(c)}) and 0.2 (\textbf{(d)} to \textbf{(f)}), at $\Omega=0.59\omega_{11}$, $\Omega=0.83\omega_{11}$ and $\Omega=1.1\omega_{11}$. An X marks the locations where particles exit the measurements domain. The wave shapes at the wall are also depicted. \textbf{(a)} \dst=0.1, starting points $z_0/H_0$=0.5, $r_0/D$=1/15, 1/5, 1/3 and 7/15.  \textbf{(b)} \dst=0.1, starting points $z_0/H_0$=2/15, 2/5 and 2/3, $r_0/D$=1/3. \textbf{(c)} \dst=0.1, $z_0/H_0$=2/15, 2/5 and 2/3, $r_0/D$=7/15. \textbf{(d)} \dst=0.2, $z_0/H_0$=0.5, $r_0/D$=1/15, 1/5, 1/3 and 7/15. \textbf{(e)} \dst=0.2, $z_0/H_0$=2/15, 2/5 and 2/3, $r_0/D$=1/3. \textbf{(f)} \dst=0.2, $z_0/H_0$=2/15, 2/5 and 2/3, $r_0/D$=7/15. \textbf{(g)} average tangential velocity $\bar{v}_\theta$ normalized by the synchronous rotation velocity $\Omega\cdot r$, for \dst=0.1 (on the left) and \dst=0.2 (on the right), the shaking frequencies are given.}
	\label{Fig_Traj} 
\end{figure*}

Concerning the liquid flow, the PIV measurements confirm the agreement between the observed flow behavior and the potential prediction of non breaking waves, while large discrepancies are observed as the wave approaches the breaking inception. The motion of the liquid is best highlighted by computing the Lagrangian trajectories followed by liquid particles released into the flow at the initial location $\mathbf{x}_0=(r_0,\theta_0, z_0)$. The positions are calculated iteratively at each successive time $t$ as follows:
\begin{align}
	& r(t+dt)  =  r(t)+v_r \big( r(t), \theta(t), z(t), t \big) \cdot dt \nonumber \\
	& \theta(t+dt)  =  \theta(t)+ \frac{1}{r(t)} v_{\theta} \big( r(t), \theta(t), z(t), t \big)  \cdot dt \label{eq_Res_LagrangianPositions}\\
	& z(t+dt)  =  z(t)+v_z \big( r(t), \theta(t), z(t), t \big)  \cdot dt,\nonumber
\end{align}
where $dt$ is the time interval, whose value (1/600 of the shaking period) is small enough not to influence the result. The values of $v_r$, $v_\theta$ and $v_z$ are estimated at each time step by linear interpolation of the velocity fields measured by PIV. The path of several liquid particles are depicted in Fig.~\ref{Fig_Traj}a for \dst=0.1 and in Fig.~\ref{Fig_Traj}d for \dst=0.2, for three shaking frequencies each. The wave patterns corresponding to the operating configurations for which the trajectories are computed are also shown.

At small shaking frequency ($\Omega=0.59\omega_{11}$) the liquid particles follow trochoidal paths, while the amplitude of their motion depends on the shaking diameter \dst~and decreases with the depth, as predicted. However, the trajectories reveal discrepancies between these potential predictions and the measurements as the shaking frequency $\Omega$ is increased: the liquid rotates in the direction of propagation of the wave. This transition is comparable to the one observed in linear forced sloshing \cite{Hutton1963}. Since the liquid is advected by the rotation, it remains in regions of overall upward or downward motion, hence giving the large vertical displacements depicted in Fig.~\ref{Fig_Traj}b, c, e and f. On the other hand, the radial motion gradually disappears. 

The evolution of the rotation motion is highlighted by averaging the tangential velocity over one revolution:
\begin{align}
	\bar{v}_\theta(r,z) = \frac{1}{2 \pi}\int_{0}^{2 \pi}{v_\theta(r,\theta,z)\,d\theta}. \label{eq_res_vThetaAverage}
\end{align}
Only the regions having a valid measurement over a complete revolution are taken into account to compute the average: i.e.~all the portions of the velocity fields that are not in the liquid phase for a part of the vessel motion do not contribute to the computation of the average. The average velocity is normalized by the wave propagation celerity at each radius: $r\Omega$, and represent therefore the fraction of synchronous rotation. The results are shown in Fig.~\ref{Fig_Traj}, for wave breaking as single (\dst=0.1) and double crested (\dst=0.2). The maximum observed rotation takes place in single crested waves slightly before the breaking inception (\dst=0.1, $\Omega=0.83\omega_{11}$), from the container revolution axis to a distance of the order of the shaking diameter \dst, whereas it diminishes approaching the external wall. The rotation is less pronounced in multiple crested waves, and the maximum is not observed at the breaking inception ($\Omega=0.692\omega_{11}$) but close to the first natural frequency. Therefore, as the shaking frequency is increased after the breaking inception, a different flow regime becomes dominant, where the inner portion of the liquid rotates rapidly around the vessel axis, the vertical motion remains significant and the radial motion has completely disappeared (as suggested by the trajectories shown in Fig.~\ref{Fig_Traj}(a) and (d)). This transition is quite abrupt at the breaking of single crested waves, while it is more gradual for breaking of multiple crested ones, as the wave shape evolves from breaking double to breaking single (see also Fig~\ref{BreakingWaves}(b)). It is interesting to notice that the maximum flow velocity at the container wall is 40-50\% of the propagation velocity of the wave, while other experiments reported 10\% for swirling waves in linear forcing \cite{Royon-Lebeaud2007} and 25\% in orbital shaking \cite{Faller2001}.

Finally, we observe that the transition to the rotational regime is not univoque as the shaking frequency is increased: we have observed that for \dst$\leq$0.02 the persistent high amplitude wave and rotating flow disappear before reaching the second natural frequency, as observed in Fig.~\ref{Fig_AmpliCompare} for \dst=0.02. This transition is subject to hysteresis: depending on the increase or decrease of the shaking frequency we may observe it at different values, as reported for sloshing due to linear forcing \cite{Royon-Lebeaud2007}.

\section{Conclusions}
We have investigated the motion of free surface liquids within orbital shaken container with the help of a potential model, wave height and velocity measurements, for a wide range of operating parameters. A large variety of wave patterns were identified, ranging from single and multiple crested waves to spilling breaking waves and liquid impinging at the wall. The potential theory correctly predicts the velocity fields and the free surface deformations for single and for specific multiple crested non-breaking waves. We have demonstrated that the occurrence of multiple crested waves is due to the excitation of sub-harmonics of particular modes. On the other hand, a generalized swirling motion of the liquid bulk is observed at the inception of the breaking, with the entire liquid rotating in the direction of propagation of the wave. While we may observe single or multiple crested waves breaking depending on the value of the shaking diameter, the maximum rotation is noticed for single crested waves at shaking frequencies slightly below the breaking inception. Since the regimes that we have illustrated affect the mixing and the oxygenation of the liquid, our research lays the foundations for the optimization of orbital shaking for different applications.

\section{Acknowledgments}
The authors would like to thank the Swiss National Science Foundation (SNSF) for the financial support through Grant No. CRSII2\_125444.

\end{document}